\documentclass[prlb,aps,12pt,a4paper,showpacs]{revtex4}
\usepackage{epsfig}
\begin{document}
\title{Symmetry-Induced Tunnelling in One-Dimensional Disordered Potentials}
\author{E.~Diez$^{1}$, F.~Izrailev$^{2}$, A.A.~Krokhin$^{3}$, A.~Rodriguez$^{1}$}
\address{$^1$Departamento de F{\'\i}sica Fundamental, Universidad de Salamanca, E-37008 Salamanca, Spain}
\address{$^2$Instituto de F{\'\i}sica, Universidad Aut\'onoma de Puebla, Apartado Postal J-48,
Puebla, 72570 Mexico}
\address{$^3$Department of Physics, University of North Texas, P.O. Box 311427, Denton, TX
76203}

\date{\today}
\begin{abstract}
A new mechanism of tunnelling at macroscopic distances is proposed
for a wave packet localized in one-dimensional disordered
potential with mirror symmetry, $V(-x)=V(x)$. Unlike quantum
tunnelling through a regular potential barrier, which occurs only
at the energies lower then the barrier height, the proposed
mechanism of tunnelling exists even for weak white-noise-like
scattering potentials. It also exists in classical circuits of
resonant contours with random resonant frequencies. The latter
property may be used as a new method of secure communication,
which does not require coding and decoding of the transmitting
signal.
\end{abstract}
\date{\today}

\maketitle

It is well-known that all quantum states in one-dimensional
white-noise potential are strongly localized and quantum transport
is limited by the distances not exceeding the localization length
$l(E)$. At longer distances the destructive interference between
direct and backscattered waves suppresses exponentially the
amplitude of a wave packet. Statistical correlations in the
disordered potential may change the interference pattern and may
give rise to a discrete set [\onlinecite{short}] or to a continuum
of delocalized states [\onlinecite{long,Izr}] for short- or
long-range correlation respectively. Correlations is a
manifestation of the local properties of a random potential. The
symmetry is a global property, therefore its effect on the
transport may be even stronger.

In this Letter we propose a symmetry-driven mechanism of
tunnelling, which is specific for the random potentials only.
Usually, the symmetry is considered to be an irrelevant property
in disordered systems since the wave functions are  localized at
the (local) scales, which are  much smaller than the (global)
scales, where the symmetry is manifested. However, the symmetry of
the potential, $V(-x)=V(x)$, leads to definite parity of the wave
functions. Either parity (even or odd) of an eigenfunction means
that there are {\it two} equal peaks with half-width $\sim l(E)$
centered at the symmetric points. A symmetry-induced correlation
between these peaks gives rise to the mechanism of tunnelling of a
wave packet (or excitation), independently how far apart the peaks
are. Due to this mechanism a wave packet tunnels at macroscopic
distances -- a process which does not exist for the random
potential without the symmetry. Natural disorder usually does not
exhibit the mirror symmetry. Nevertheless, the proposed mechanism
of tunnelling is not of pure academic interest, since it may be
observed also in a classical system -- a random electrical
circuit, where the symmetry can be easily introduced. In what
follows we propose a new method of secure communications based on
the symmetry-induced mechanism of tunnelling. The merit of this
method is that it does not require a coding-decoding procedure.

To demonstrate the main idea of the symmetry-induced tunnelling we
consider the tight-binding Anderson model [\onlinecite{And}].  For
one-dimensional lattice a stationary solution for the eigenstate
with energy $E$ is obtained from the equation
\begin{equation}
\label{tight} \psi _{n+1}+\psi _{n-1}=(E+\epsilon _n)\,\psi _n \,,
\end{equation}
where $\epsilon _n$ is on-site energy. The energies $E$ and
$\epsilon _n$ are measured in units of the hopping amplitude $t$,
which in the case of diagonal disorder is independent on the site
index $n$.

\begin{figure}
\includegraphics[width = 8cm]{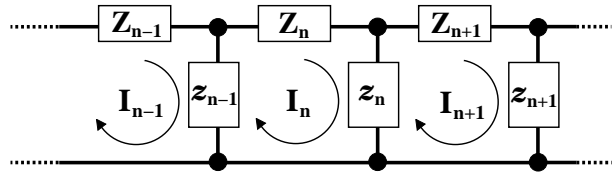}
\caption{Segment of infinite electric circuit of classical
impedances.} \label{fig1}
\end{figure}

Discrete Schrodinger equation (\ref{tight}) gives {\it exact}
description  of the electrical circuit of classical impedances
$Z_n$ and $z_n$ shown in Fig. {\ref{fig1}}. Application of
Kirchhoff's Loop Rule to three successive unit cells of the
circuit leads to the following linear relation between the
currents circulating in the $(n-1)$-th, $n$-th and $(n+1)$-th
cells
\begin{equation}
\label{imped} z_nI_{n+1} + z_{n-1}I_{n-1} = (Z_n + z_n +
z_{n+1})I_n \,.
\end{equation}
If the vertical impedances are all the same, $z_n = z_0$, this
equation is reduced to the tight-binding model with diagonal
disorder, Eq. (\ref{tight}), with $\epsilon_n = \delta_n/z_0$ and
$E = 2 + Z_0/z_0$. Here the random value of the impedance $Z_n$ is
split into its mean value $Z_0 = \langle Z_n \rangle$ and the
fluctuating part $\delta_n = Z_n - Z_0$.

This exact correspondence allows testing of quantum effects of
Anderson localization using classical electrical circuits with
random elements. In fact, during the last decade chaotic resonant
cavities have been successfully used for testing the predictions
of quantum chaos [\onlinecite{Sto}].
 It is worth mentioning that electrical circuits of passive elements
 have been widely used for modelling different physical phenomena.
 The first application of the method of equivalent
 circuit probably goes back to Lord Kelvin who used a discrete $RC$
 chain to study a signal transmission through a transatlantic cable.
 Many bright examples of electrical circuits that model quantum mechanical behavior for
 simple but fundamental systems are given in the book by Pippard
 [\onlinecite{Pip}]. Recently it was proposed that electromagnetic
 waveguide can be used to model as exotic effect as Hawking black hole radiation
 [\onlinecite{Haw}]. Some effects of correlated disorder have been studied
 in the experiments with microwave propagation through disordered
 waveguides [\onlinecite{Uhl}] and sub-terahertz response of superconducting multilayers
 [\onlinecite{super}].
Experimental realization of a system with desirable correlations
and observation of the localized and extended states are much
easier in electromagnetic devices [\onlinecite{Uhl,super}] than in
heterostructures with intentionally introduced disorder
[\onlinecite{Bel}].

If the potential in Eq. (\ref{tight}) is an even function,
$\epsilon_n = \epsilon_{-n}$, the eigenfunctions
${\Psi}^{\alpha}_n$ are either even or odd functions of $n$. If an
eigenfunction ${\Psi}^{\alpha}_n$ is localized near a site $n_0$,
the amplitude of this state at the origin is exponentially
suppressed, $\Psi^{\alpha}_{n=0} \propto \exp(-\mid n_0
\mid/l(E_{\alpha}))$, provided $\mid n_0 \mid \gg l(E_{\alpha})$.
However, due to definite parity of the wave function, another peak
appears at the symmetric point $n=-n_0$. Strong localization of
any excitation in random potential is a result of {\it
destructive} interference between propagating and backscattered
wave. The appearance of the symmetric peak can be explained as a
result of {\it constructive} interference. It leads to exponential
increase of the amplitude of the wave, i.e. to {\it
antilocalization} [\onlinecite{symmetry}].

\begin{figure}
\includegraphics[width = 8cm]{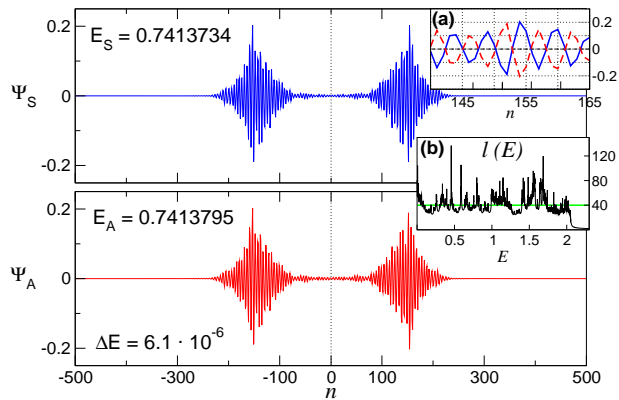}
\caption{(Color on line) Two eigenstates with different parity
($\Psi_S$ is even and $\Psi_A$ is odd) in random symmetric
potential $\epsilon_{-n} = \epsilon_n$ with $\langle \epsilon_n
\rangle=0$ and $\langle \epsilon_n^2 \rangle = \epsilon_0^2=0.1$.
These states belong to a doublet with energy splitting $\Delta E$.
Inserts: (a) Blow-up of the right peaks of the eigenfunctions
showing that they possess different parity. (b) Numerical result
for the localization length as compared to the energy-independent
function $l(E)=40$. The compensation of the energy dependence in
$l_0(E)$ is not of principal importance and is done only to
simplify the discussion of the numerical results.} \label{fig2}
\end{figure}

In Fig. {\ref{fig2}} we show two quasi-degenerate eigenstates
calculated for the  symmetric potential of 1000 sites (i.e. only
500 of these sites are random).
The inverse localization length (the Lyapunov exponent) can be
estimated from the formula [\onlinecite{Izr}]
\begin{equation}
\label{loclength} l^{-1}(E)= l_0^{-1}(E)\varphi(\mu),\,\,
\varphi(\mu )=1+2\sum\limits_{k=1}^\infty \xi (k)\,\cos \,(2\mu
\,k).
\end{equation}
Here $l^{-1}_0 (E)= \epsilon _0^2/ (8\sin^2\mu )$ is the Thouless
[\onlinecite{Tho}] result for the white noise disorder, the
function $\varphi(\mu)$ accounts for the contribution of
correlations with correlation function $\langle\epsilon_n
\epsilon_{n+k}\rangle = \epsilon_0^2 \xi(k)$, and the dispersion
relation is $E=2\cos\mu$. The results shown in Fig. {\ref{fig2}}
are obtained not for white-noise but for slightly correlated
disorder with correlator $\xi(1) = -1/2$ and $\xi(k>1)=0$. These
short-range correlations are introduced in order to compensate the
smooth energy dependence of $l_0(E)$. It is easy to see that the
contribution of the term with $k=1$ in Eq. (\ref{loclength})
provides a flat dependence $l^{-1}(E)=\epsilon_0^2/4=const$.
Insert (b) in Fig. \ref{fig2} shows the numerical values of
$l(E)$, which fluctuate around $40$ -- the value obtained from Eq.
(\ref{loclength}). In agrement with this estimate, the half-width
of the peaks in Fig. \ref{fig2} is approximately $40$ sites.

The energy spectrum of Eq. (\ref{tight}) with symmetric random
potential is similar to the spectrum of a double-well potential.
It consists of discrete levels, most of them lying within the
interval $-2<E<2$. The energy levels are arranged in doublets of
states with different parity. The energy $\delta (E)$ between the
centers of the doublets scales with the length of the system $N$
as $1/N$. The energy splitting $\Delta E $ in the doublet is
exponentially small, $\Delta (E) \propto \exp[-4\mid n_0
\mid/l(E)]$, i.e. the states are quasi-degenerate.  Both, $\delta
(E)$ and $\Delta(E)$ fluctuate with energy because of statistical
fluctuations of the density of states, $n_0(E)$, and $l(E)$.

The symmetry-induced tunnelling can be observed in the dynamics of
an excitation. Let a perturbation is applied at one of the sites
of the symmetric random sequence. In the simplest case the
perturbation is a $\delta$-excitation at the site $n_0$,
$\psi_n(t=0) = \delta_{nn_0}$. Since this excitation is not an
eigenfunction of the system, its temporal evolution is represented
as a superposition,
\begin{equation}
\label{initial} \psi_n(t)=\sum_{\alpha} C^{\alpha}_{n_0}
\Psi_n^{\alpha}\exp(-iE_{\alpha}t).
\end{equation}
The sum in Eq. (\ref{initial}) runs over the eigenstates, which
are all localized. The eigenstates centered closer to the initial
excitation contribute more because the coefficient
$C^{\alpha}_{n_0} = \langle \Psi_n^{\alpha}|\psi_n(t=0)
 \rangle = \Psi_{n_0}^{\alpha}$ is the overlapping
integral between the initial excitation and the eigenstate
$\Psi_n^{\alpha}$. Let the eigenstates with maximum overlapping be
$\Psi_A$ and $\Psi_S$. They form a doublet with the central energy
$\bar E =(E_A+E_S)/2$ and splitting $\Delta E = E_A - E_S$. Taking
into account only these two terms in Eq. (\ref{initial}), the
following approximate result for the evolution of the initial
excitation can be easily obtained
\begin{equation}
\label{evolution} \psi_n(t) \approx \frac{e^{{-i{\bar
E}t}}}{\sqrt{l(\bar{E})}} \left[ \cos\left(\frac{\Delta E
}{2}t\right) \Psi_+(n) + i \sin\left(\frac{\Delta E }{2}t\right)
\Psi_-(n) \right] .
\end{equation}
Here $\Psi_{\pm }(n)=(\Psi_S {\pm} \Psi_A)/\sqrt 2$. Each of these
linear combinations is a single-peak function. The peak of
$\Psi_+$ is always close to the point of initial excitation. For
the eigenfunctions shown in Fig. {\ref{fig2}} the peak $\Psi_+$ is
localized in the region of negative $n$.

At the early stage of evolution the initial $\delta$-peak at $n_0$
quickly spreads over the region of width $2l(\bar{E})$. Further
spreading is suppressed by Anderson localization. Eq.
(\ref{evolution}) is not valid at this transient stage. This
equation describes steady and ``slow" harmonic oscillations of the
initial excitation between the two symmetrical points. The period
of oscillations of the density $\mid \psi_n(t) \mid^2$ is
$T=2\pi/\Delta E$. If the distance $2\mid n_0 \mid$ between the
peaks exceeds the localization length, the amplitude of the wave
function at the origin is exponentially small $\sim \exp(-\mid n_0
\mid/l(\bar{E}))$. It, however, grows exponentially towards the
symmetrical point $-n_0$. This increase is a manifestation of the
tunnelling induced by the symmetry. The dynamics of penetration of
the initial excitation to the symmetrical point is very similar to
the tunnelling through a potential barrier, although there is no
real barrier. Exponential decrease (increase) of the wave
functions is due to multiple scattering events with predominant
destructive (constructive) interference. One can speak about an
effective double-well potential which produces the same discrete
energy spectrum. Calculation of the parameters of this effective
potential is a challenging inverse-scattering problem. Tunnelling
processes without a real barrier are known in dynamical systems,
where quantum transitions occur either between strongly localized
states [\onlinecite{Cas}] or between classically separated regions
in phase space [\onlinecite{Pod}]. It is worth mentioning that
regular Bloch-like oscillations in a potential with correlated
disorder may occur also due to the presence of two mobility edges
in the energy spectrum [\onlinecite{Adam}].
\begin{figure}
\includegraphics[width = 8cm]{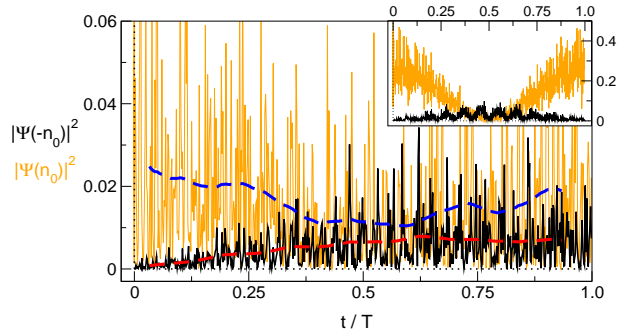}
\caption{(Color on line) Temporal evolution of the probability
density $\mid \psi(t) \mid^2$ calculated at the point of
excitation, $n_0= -153$ (grey line) and at the symmetric point,
$-n_0=153$ (black line). Dashed lines are the window-average
values of the densities at the symmetric points and are drawn for
guide-eye only. The insert shows similar temporal evolution but
for much stronger disorder, $\epsilon_0^2 = 0.55$. In this case
the localization length is shorter $l(E) = 8$ and much less
doublets contribute to the evolution of the the initial
excitation.}\label{fig3}
\end{figure}

Eq. (\ref{evolution}) takes into account interaction of the
initial excitation with the nearest doublet. If there are more
eigenstates in Eq. (\ref{initial}), whose wave functions extend to
the point $n_0$, they also contribute to the evolution of the
initial excitation. In this case the oscillations between the
peaks at $n_0$ and $-n_0$ are not harmonic any more but a
superposition of harmonics with different periods. In the
numerical study of evolution of the excitation we take into
account its interaction with the eigenfunctions which have
amplitude $> 10^{-3}$ at the site $n_0= -153$.  There are 540 such
eigenfunctions out of total 1000. These eigenfunctions produce the
oscillatory pattern in Fig. \ref{fig3}. Although the site
$n_0=-153$ is the position of the maximum for the eigenstates
$\Psi_A$ and $\Psi_S$ in Fig. {\ref{fig2}}, other states give a
noticeable contribution. Since the levels splittings in different
doublets are random and incommensurate, the evolution of the wave
packet is not periodic but it keeps the main features predicted by
Eq. (\ref{evolution}). In the case of stronger localization of the
eigenstates the dependence $\mid \psi_{n_0}(t) \mid^2$ approaches
the harmonic dependence Eq.(\ref{evolution}) as it is seen in the
insert in Fig. \ref{fig3}.

\begin{figure}
\includegraphics[width = 8cm]{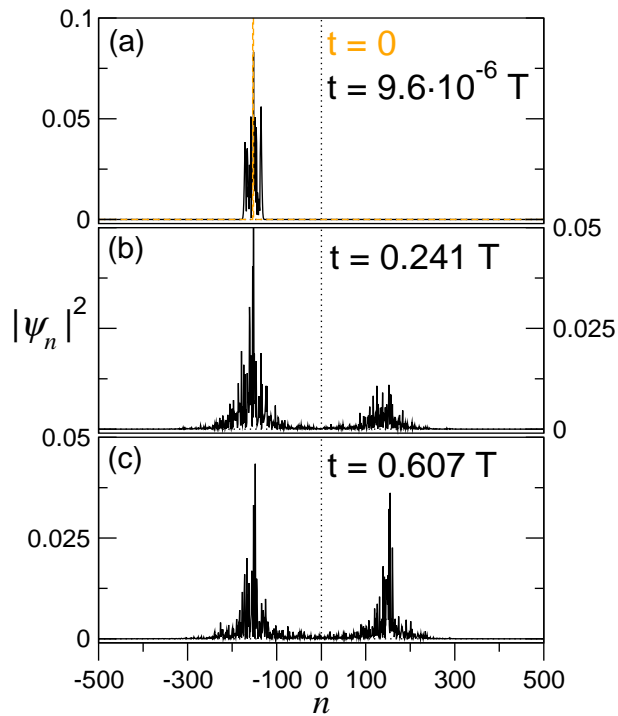}
\caption{(Color on line) Spatial distribution of the probability
$\mid \psi_n(t) \mid^2$ at three instants: (a) Spreading of the
initial peak at the transient stage, $t=9.6\cdot 10^{-6}T$. The
initial peak of amplitude 1 at $n_0=-153$ is shown in grey. The
secondary peak is not visible at this stage. (b) The secondary
peak at $-n_0=153$ is well developed at $t = 0.241 T$; (c) The two
peaks become almost equal at $t =0.607 T$. }\label{fig4}
\end{figure}

Spreading and tunnelling of the initial excitation is shown in
Fig. {\ref{fig4}}. At the transient stage the excitation broadens
up to the size of $2l(\bar E)\approx 80$, Fig. \ref{fig4}a. The
initial stage is followed by the long-lasting stage of tunnelling
at the macroscopic distance $2\mid n_0 \mid$. The tunnelling gives
rise to the secondary peak at $-n_0$, which "slowly" grows and
reaches its maximum at $t \approx T/2$, Fig. \ref{fig4}c. The
amplitude of the secondary peak in Figs. {\ref{fig3}} and
\ref{fig4}c does not exceed $5\%$ of the initial peak, but at $t
\approx T/2$ the both peaks have approximately the same amplitude.
The amplitude of the peaks can be obtained from the normalization
condition and it is determined by $l(E)$ as it follows from Eq.
(\ref{evolution}). This amplitude is much larger then the
amplitude at the origin $n=0$, as it can be clearly seen from Fig.
\ref{fig4}. The amplitude of the peaks at the points $\pm n_0$
increases with $\epsilon_0$. Simultaneously the wave function at
the origin decreases exponentially and can be easily controlled by
the disorder.

Application of the proposed ideas to random electrical circuits is
straightforward. In what follows we demonstrate the evolution of
the signal in a circuit shown in Fig. {\ref{fig1}} with vertical
impedances being equal solenoids with $z_n= -i\omega L_0$ and
horizontal impedances being capacitors with $Z_n = \frac{i}{\omega
C_n}  \approx \frac{i}{\omega C_0}\left(1- \frac{\delta C_n}{C_0}
\right) $. Here $\delta C_n$ is the fluctuating part of the
capacitance, which is an even function of $n$, $\delta C_n =
\delta C_{-n}$. Propagation of an excitation in this circuit
follows the wave equation
\begin{equation}
\label{waveeq} C_{n+1}\ddot{V}_{n+1} + C_{n-1}\ddot{V}_{n-1} -2C_n
\ddot{V}_n = V_n/L_0,
\end{equation}
where $V_n$ is voltage drop at the $n$-th capacitor. This equation
requires two initial conditions. Let the voltage drop $V_0$ is
applied at $t=0$ to the capacitor $C_{n_0}$, inducing the initial
current $I_0=C_{n_0}\dot V_0$. Stationary solutions ($V_n \propto
\exp (-i\omega t)$) are either even or odd functions and the
spectrum of eigenfrequencies consists of a set of doublets. For an
infinite chain the majority of the eigenfrequencies occupy an
interval $[\omega_0/2,\infty]$, where $\omega_0=\left(L_0C_0
\right)^{-1/2}$. Assuming that the initial perturbation excites
only the closest to the site $n_0$ pair of eigenstates ($V_A$ and
$V_S)$, the solution of Eq. (\ref{waveeq}) can be written in the
form similar to Eq. (\ref{evolution}),
\begin{eqnarray}
\label{evolutioncir} V_n(t) \approx \sqrt{\frac{1}{C_0
l(\bar\omega)}} \left\{\left[C_{n_0} V_0 \sin(\bar{\omega}t)-
\frac{I_0}{\bar{\omega}} \cos(\bar{\omega}t) \right]\sin\left(
\frac{\Delta \omega}{2}t\right) V_-(n) \right.\nonumber \\
+\left. \left[C_{n_0} V_0 \cos(\bar{\omega}t)+
\frac{I_0}{\bar{\omega}} \sin(\bar{\omega}t) \right] \cos\left(
\frac{\Delta \omega}{2}t\right) V_+(n) \right\}.
\end{eqnarray}
Here $\bar \omega$ is the center of the doublet, and $\Delta
\omega$ is the frequency splitting. Single-peak functions
$V_{\pm}(n)= (V_S \pm V_A)/\sqrt2$  play the same role as
$\Psi_{\pm}$ do in Eq. (\ref{evolution}). Eq. (\ref{evolutioncir})
shows that the evolution of the initial signal in a random
(symmetric) electrical circuit is similar to the wave packet
evolution obtained from the tight-binding model. There is an
obvious symmetry-induced tunnelling of the initial signal at
macroscopic distances.

The effect of tunnelling can be used for secure communications.
Instead of coding and decoding a signal, we propose to suppress
the transmitted signal by a circuit with random elements and then
to restore it, using a symmetric counterpart of the random
circuit. The signal can be suppressed to the noise level and
safely transmitted to the receiver over a transmitting line. The
symmetric counterpart of the random circuit restores only the
signal, (not noise) since constructive interference occurs only
for the coherent part, which has passed trough the suppressing
circuit of the emitter. The non-coherent part (noise or any
irrelevant signal) will be exponentially suppressed by the
receiving random circuit. The two identical random circuits may be
fabricated as microchips, which are installed (or replaced)
simultaneously at the emitter and receiver. In the absence of
dissipation and asymmetry between the two random elements, the
proposed method guarantees a robust restoration of the signal.
Inevitable Joule losses should be compensated by an amplifier,
which does not destroy the coherency of the signal.

In conclusion, we demonstrate that in a symmetric random potential
the localized quantum states have two peaks as it is required by
the parity. Fast spreading of the initial $\delta$-excitation
within localization length is followed by slow growth due to
tunnelling at the symmetrical point. This  effect opens a new
possibility for secure communications that does not require coding
and decoding of the transmitting signal. The random circuits may
operate in a wide range of radio-frequencies, using commercial
capacitors and inductors. They also can be fabricated and operated
in the infrared and optical region using the concept of plasmonic
nanoelements proposed in Ref.[\onlinecite{opt}].

This work is supported by the US DOE grant $\#$ DE-FG02-06ER46312,
MMA (Spain) grant $\#$ A106/2007 , MEC (Spain) grant $\#$
FIS2006-00716, and by CONACyT (Mexico) grant $\#$ 43730. E.D
acknowledges support from Ram\'on y Cajal fellowship.


\begin{thebibliography}{99}
\bibitem{symmetry} In a symmetric random potential the correlations
always exist for the symmetric points, $\langle\epsilon_n
\epsilon_{-n} \rangle=\epsilon_0^2$. At the same time the nearest
sites may be uncorrelated. Because of this property the
correlation function is not stationary. But it is stationary if
the both sites are on the same side from the origin,
$\langle\epsilon_n \epsilon_{n^{\prime}} \rangle= \epsilon_0^2
\xi(n-n^{\prime})$, if $nn^{\prime}>0$.
\end{thebibliography}
\end{document}